\documentclass[preprint,12pt]{elsarticle}

\usepackage{graphicx}
\usepackage{amsmath,amssymb}
\usepackage{xcolor}
\usepackage{float}
\usepackage{array}
\usepackage{adjustbox}
\usepackage{booktabs}
\usepackage{lineno}
\usepackage{algorithm}
\usepackage{algpseudocode}
\usepackage{subfig}

\begin{document}
\begin{frontmatter}

\title{How Exploration Breaks Cooperation in Shared-Policy Multi-Agent Reinforcement Learning}

\author[addr1]{Yi-Ning Weng}
\author[addr2]{Hsuan-Wei Lee\corref{cor1}}
\ead{hsl324@lehigh.edu}
\cortext[cor1]{Corresponding author.}
\address[addr1]{Department of Accounting, National Taiwan University, Taiwan}
\address[addr2]{College of Health, Lehigh University, USA}

\date{}

\begin{abstract}

Multi-agent reinforcement learning (MARL) in dynamic games poses fundamental challenges due to non-stationary interactions and exploration noise, which are often compounded by common design choices such as parameter sharing. While parameter sharing is commonly adopted to improve sample efficiency and scalability, its implications for cooperative behavior under exploration remain poorly understood. In this work, we conduct a systematic study of Deep Q-Networks (DQN) with parameter sharing in a dynamic Prisoner’s Dilemma environment, focusing on how exploration and payoff harshness jointly shape learning stability and cooperation outcomes.

We reveal a robust cooperation collapse in shared DQN as exploration increases. The collapse is most severe when payoffs strongly favor defection. This collapse is not sporadic but forms structured failure regions in the parameter space, with structured failure boundaries that shift systematically with game severity. This phenomenon cannot be explained by payoff incentives or learning instability alone. Cooperation collapse coincides with degradation in learned representations: cooperative and defective behaviors become entangled in the feature space, and action-value gaps shrink systematically. The system converges to stable but low-cooperation solutions.

We identify partial observability and representational coupling as key drivers. Grouped policy learning, which limits parameter sharing to subsets of agents, consistently alleviates collapse. State augmentation with learning progress and exploration signals significantly delays collapse and partially reverses it in some regimes. Network topology also matters: spatial structure supports cooperation through local coordination, while random networks eliminate cooperative regimes. These interventions confirm that the failure is representational, not algorithmic.

Our findings reveal a fundamental failure mode of parameter-shared Deep Q-Networks in cooperative settings: shared representations cannot maintain stable cooperation under exploration. However, representation-aware state design can substantially improve robustness. These results have implications for designing scalable MARL systems where cooperation is essential.

\end{abstract}


\end{frontmatter}

\section{Introduction}

Multi-agent reinforcement learning (MARL) provides a framework for modeling systems in which multiple autonomous agents learn through repeated interaction. Applications range from distributed control and networked systems to large-scale coordination problems, where individual decision-makers must adapt to both environmental feedback and the evolving behaviors of others. Exploration poses a fundamental challenge in reinforcement learning. This challenge intensifies in multi-agent settings, where concurrent learning agents induce non-stationarity in the effective environment each agent experiences~\cite{hong2018diversity, shen2019multi, hao2024exploration_survey}. While structured exploration mechanisms have been proposed to improve sampling efficiency in single-agent DQN~\cite{huang2024preference_guided_exploration}, their interaction with parameter sharing in multi-agent settings remains poorly understood. Understanding how learning architectures interact with non-stationarity and exploration is therefore essential for building scalable and robust MARL systems \cite{hernandez2017survey, marinescu2017prediction, xie2021deep, steinparz2022reactive, liu2025learning, carvalho2025reinforcement}.

Parameter sharing has become a standard approach to address scalability in MARL \cite{christianos2021scaling, li2024adaptive, zhang2024scalable}. By training multiple agents using a shared policy or value function, this design reduces sample complexity and enables learning in systems with large agent populations. Many value-based cooperative MARL methods adopt centralized training with decentralized execution, incorporating information about teammates' behaviors during training~\cite{liu2024_teammates_actions, pei2025policy}. Despite these advantages, parameter sharing imposes a fundamental constraint: all agents are coupled through a single representational and optimization pathway \cite{lee2020parameter, fumero2023leveraging}. Every agent contributes gradients to, and draws decisions from, the same network, regardless of differences in local interaction contexts. As a result, shared policies are exposed to aggregated exploration noise and non-stationary feedback from many interacting agents \cite{padakandla2020reinforcement}. Parameter sharing is often treated as a benign engineering choice. However, its implications for learning stability and cooperation under exploration remain poorly understood, particularly when agents face heterogeneous and evolving interaction contexts.

This gap in understanding is particularly critical in cooperative settings. Sustained coordination requires stable feedback between individual actions and collective outcomes. Small perturbations from exploration or miscoordination can propagate through interactions and disrupt cooperative structures \cite{kapetanakis2003reinforcement, li2020adaptive, chen2024agentverse}. Cooperation breakdowns are frequently observed in shared-policy MARL under exploration. However, they are rarely studied as systematic phenomena. Three potential explanations exist: unavoidable incentive conflicts, training instability, or fundamental limitations in how shared representations handle non-stationary multi-agent interactions. This work distinguishes among these possibilities. Network structure also plays a role in cooperative learning. Prior work shows that network topology and adaptive rewiring influence the emergence and stability of cooperation in learning systems \cite{pal2022network, lee2025enhancing, weng2025q, santos2006graph, kinne2013network, rand2014static, fotouhi2019evolution, shiu2025shaping}. We examine how interaction structure modulates the cooperation collapse we identify.

We investigate these questions through a systematic study of parameter-shared Deep Q-Networks in a dynamic Prisoner's Dilemma. This environment serves as a diagnostic setting that allows controlled manipulation of incentive structure, exploration strength, and network topology \cite{nowak1992evolutionary, szabo2007evolutionary, yu2015emotional, mengel2018risk, lee2018evolutionary}. We use cooperation rate not as an optimization objective, but as an indicator of learning stability and coordination quality. Our main finding is a robust cooperation collapse in parameter-shared Deep Q-Networks as exploration increases. This collapse occurs across broad regions of the parameter space and forms structured failure boundaries that shift systematically with game severity. 

To understand this failure, we examine the internal dynamics of the shared network. We find that cooperation collapse coincides with degradation in learned representations. Cooperative and defective behaviors become increasingly entangled in the latent space. At the value level, action-value gaps contract systematically. This suggests convergence to stable but low-cooperation attractors, not training instability. These findings rule out two simple explanations: incentive misalignment and excessive exploration noise. Instead, the failure is rooted in how shared representations handle non-stationary multi-agent learning.

If representational coupling causes the failure, then reducing it should help. We test this prediction using grouped policy learning. Grouped policy learning limits parameter sharing to subsets of agents. This consistently improves stability and delays collapse, confirming that global representation sharing contributes to the failure. We next test whether the failure stems from partial observability of learning dynamics. We augment the state with signals encoding learning progress and exploration intensity. This augmentation delays cooperation collapse across all exploration levels. In some regimes, it partially reverses the collapse, restoring stable cooperation over a wider parameter range. These interventions show that state and representation design are critical for shared-policy MARL. Exploration tuning alone cannot achieve robustness. We also examine how network topology affects cooperation collapse. Spatially structured grids support cooperation by providing local coordination scaffolding. Random topologies eliminate this benefit, making cooperation collapse more severe. This reveals a boundary condition: parameter-shared DQN requires environmental structure to sustain cooperation under exploration. Learning dynamics alone are insufficient.

This work identifies and characterizes a fundamental failure mode of parameter-shared Deep Q-Networks in cooperative settings. We show that representation-aware state design can mitigate this failure. Our findings link macro-level cooperation failure to internal representation dynamics. This provides guidance for designing more robust shared-policy MARL systems.

\section{Methods}

We study a multi-agent reinforcement learning system in which agents learn to act in a dynamic Prisoner’s Dilemma Game (PDG) using Deep Q-Networks (DQN). The model builds upon established frameworks that integrate reinforcement learning with evolutionary game dynamics, and is designed to examine learning behavior under non-stationary interactions induced by simultaneous adaptation and exploration\cite{padakandla2020nonstationary}.

A population of agents is embedded in a fixed interaction network and repeatedly engages in local pairwise games with neighboring agents. At each discrete time step, every agent selects one of two available actions: cooperation ($C$) or defection ($D$). Interactions are symmetric and occur concurrently across the network. Following each round of interactions, agents receive payoffs determined by the local game outcomes and update their decision policies based on accumulated experience. This setting induces an endogenously non-stationary learning environment, as the effective dynamics faced by each agent evolve over time in response to the concurrent learning and exploration of others. The following subsections formally specify the dynamic game environment, the shared-policy learning architecture, the agents’ state representation and its partial observability, the exploration mechanism, and the learning procedure used throughout this study. Throughout this work, cooperation ($C$) is encoded as 0 and defection ($D$) is encoded as 1 in the binary state representation.

\subsection{Prisoner's Dilemma Payoff Structure and Interaction Network}

We use a dilemma-strength parameterization that allows independent control of exploitation risk and temptation to defect \citep{LeeWeng2025GranularQ,weng2025q,geng2022reinforcement,wang2013spatial}. The payoff matrix is defined as
\begin{equation}
\label{eq:pd_payoff}
\begin{pmatrix}
R & S \\
T & P
\end{pmatrix}
=
\begin{pmatrix}
1 & -D_r \\
1 + D_g & 0
\end{pmatrix},
\end{equation}
where $R=1$ is the reward for mutual cooperation, $P=0$ is the punishment for mutual defection, $T=1+D_g$ is the temptation to defect, and $S=-D_r$ is the sucker's payoff for being exploited.

The parameters $D_g = T - R$ and $D_r = P - S$ control two aspects of dilemma strength. Both range from 0 to 1. Specifically, $D_g \in [0,1]$ captures the temptation advantage of unilateral defection, corresponding to the chicken-type dilemma component, whereas $D_r \in [0,1]$ measures the punishment severity associated with being exploited, reflecting the stag-hunt-type dilemma element. This parameterization covers the full space of two-player social dilemmas when $(D_r, D_g)$ varies over $[0,1]^2$.

Within this framework, $D_r$ primarily controls the risk of exploitation, while $D_g$ modulates the reward incentive for defection. Different regions of the $(D_r, D_g)$ plane correspond to distinct classes of social dilemmas: when $D_r > D_g$, the system exhibits stag-hunt-like characteristics emphasizing coordination benefits, whereas $D_g > D_r$ yields chicken-game-like dynamics dominated by temptation to defect. Consistent with prior findings that cooperation dynamics are more sensitive to changes in $D_r$ than in $D_g$ \cite{tanimoto2007dilemma, wang2015universal}, we focus our analysis on the diagonal constraint $D_g = D_r$, and systematically vary $D_r$ within a restricted range.

Agents are arranged on a two-dimensional square lattice of size $L \times L$, giving $N = L^2$ agents total. We use periodic boundary conditions. Each agent interacts with its four nearest neighbors (von Neumann neighborhood). The network structure remains fixed during each run. This spatial structure induces repeated local interactions while limiting long-range information propagation, providing a controlled setting for studying cooperative learning under local interaction constraints. Spatial clustering and local interaction patterns on networks have been shown to substantially affect the evolution of cooperation in game-theoretic settings \cite{kuperman2012relationship, lee2022costly, si2025evolution}. At each time step, every agent plays the game simultaneously with all four neighbors. Each agent's reward is the mean payoff from these four interactions. The environment evolves in discrete time steps. At each step, agents observe local states, select actions concurrently, receive rewards, and update their policies. The network topology and payoff matrix remain fixed.

In addition to the grid network, we also test three alternative topologies: modular 4-regular, random 4-regular, and small-world networks. All topologies maintain an average degree of four to ensure fair comparison. Unless otherwise stated, the payoff structure, state representation, learning algorithm, and training protocol are kept identical across different topologies, with only the interaction graph varied. The effects of interaction topology on cooperation outcomes are examined in the Results section.

\subsection{Shared-Policy Deep Q-Network Architecture}

All agents share a single Deep Q-Network (DQN). One action-value function is used by the entire population for both learning and action selection. We refer to this architecture as the \emph{Shared DQN} throughout the remainder of the paper. Parameter sharing is adopted to promote scalability and sample efficiency, such that all agents map their local observations to action-value estimates through a common function approximator.

Let $s_i^t$ denote the local state observed by agent $i$ at time step $t$, and let $a \in \{C,D\}$ denote the available actions. The shared DQN approximates a common action-value function
\begin{equation}
Q_\theta(s_i^t,a),
\end{equation}
parameterized by $\theta$, which is used by all agents for action selection. Agents experience different local contexts, but all use the same shared network for decisions and learning.

The network has one hidden layer with 96 ReLU units and a linear output layer. The output layer produces Q-values for both actions (cooperate and defect). A separate target network $Q_{\theta^-}$ with identical architecture is maintained to stabilize training, and its parameters are periodically synchronized with those of the online network. At each time step, all agents generate experience tuples $(s, a, r, s')$. These experiences from all agents go into a single shared replay buffer. Mini-batches are sampled uniformly from this buffer, mixing experiences from all agents. Gradient updates therefore aggregate signals from the entire population, coupling all agents through one shared network.

The shared DQN is trained using a temporal-difference objective. For a transition $(s_t, a_t, r_t, s_{t+1})$, the one-step target is defined as
\begin{equation}
y_t^{(1)} = r_t + \gamma \, Q_{\theta^-}\!\left(s_{t+1}, \arg\max_{a'} Q_\theta(s_{t+1}, a')\right),
\end{equation}
where $\gamma$ is the discount factor, $y_t^{(1)}$ denotes the one-step temporal-difference target, and $a'$ indexes the available actions at the next state. This formulation follows the Double DQN principle, in which action selection and action evaluation are decoupled between the online and target networks to reduce overestimation bias.

To incorporate longer-term credit assignment, we additionally employ an $n$-step return. When a valid sequence of transitions from the same agent is available, the $n$-step target is given by
\begin{equation}
y_t^{(n)} = \sum_{k=0}^{n-1} \gamma^k r_{t+k} + \gamma^n Q_{\theta^-}\!\left(s_{t+n}, \arg\max_{a'} Q_\theta(s_{t+n}, a')\right).
\end{equation}
In our implementation, we employ an $n$-step return with $n=5$ whenever such sequences can be reliably constructed, and fall back to one-step updates otherwise. This hybrid strategy ensures consistent learning under asynchronous experience collection.

The network parameters are optimized using the AdamW optimizer with weight decay \cite{llugsi2021comparison, reyad2023modified}. The loss function is defined as the Huber loss between predicted and target action values, providing robustness to outliers in temporal-difference errors. Gradients are clipped to a fixed norm to improve numerical stability during training.

We also consider a grouped variant as a diagnostic baseline. In this variant, agents are partitioned into 10 disjoint groups. Each group maintains its own shared DQN with identical architecture and learning procedure. Agents are assigned to groups randomly (not by spatial location). We refer to this baseline as the \emph{Grouped DQN}. Parameters are shared among agents within the same group but not across different groups, while all other components of the environment and training process are kept identical. This is a diagnostic tool to test whether global representation sharing causes the failures we observe.

Together, these architectures define fully decentralized execution schemes with centralized learning at the group or population level, with the fully shared-policy DQN serving as the primary learning architecture analyzed throughout this study.

\subsection{State Representation and Partial Observability}

Each agent makes decisions based on a local state representation that encodes recent interaction information with its immediate neighbors. Specifically, the baseline state of agent $i$ at time step $t$ is defined as a fixed-length binary vector
\begin{equation}
s_i^t = \big( a_{j_1}^{t-1}, a_{j_2}^{t-1}, a_{j_3}^{t-1}, a_{j_4}^{t-1}, a_i^{t-1} \big),
\end{equation}
where $j_1,\dots,j_4$ denote the four nearest neighbors of agent $i$ on the interaction network, and $a \in \{0,1\}$ indicates the action taken in the previous time step, corresponding to cooperation or defection. This state includes only the last timestep. It contains no history beyond $t-1$, no global information, and no learning progress signals.

This minimal state is standard in spatial evolutionary games. However, it induces partial observability when agents learn concurrently. State transitions depend on neighbors' policies, which change during training. The environment is therefore non-stationary from each agent's perspective. Importantly, the state representation does not encode any information about the stage of training, exploration intensity, or global learning dynamics. Consequently, agents operating under a shared policy must infer appropriate actions solely from local interaction signals, even as the underlying population-level behavior and exploration regime change over time.

In later experiments, we systematically augment this baseline state with additional low-dimensional signals that expose latent temporal or learning-progress information to the shared representation. These augmented states are constructed by appending scalar features to the original interaction-based state while keeping the learning architecture, optimization procedure, and environmental dynamics unchanged. This design allows us to isolate the effect of state observability on learning outcomes without introducing additional algorithmic modifications.

\subsection{Exploration Mechanism}

Action selection is governed by a stochastic exploration mechanism based on a softmax (Boltzmann) policy. Given the action-value estimates produced by the shared DQN, each agent selects actions probabilistically according to
\begin{equation}
\pi(a \mid s) = \frac{\exp\!\left(Q_\theta(s,a)/\tau\right)}{\sum_{a'} \exp\!\left(Q_\theta(s,a')/\tau\right)},
\label{eq:softmax_policy}
\end{equation}
where $\tau > 0$ denotes the temperature parameter controlling the strength of exploration. For numerical stability, we apply the standard logit-shifting trick by subtracting $\max_{a} Q_\theta(s,a)$ before computing the softmax, which leaves $\pi(a\mid s)$ unchanged. Larger values of $\tau$ induce more uniform action sampling, while smaller values concentrate probability mass on higher-valued actions.

The temperature parameter $\tau$ is annealed over training according to a fixed schedule. Specifically, $\tau$ is initialized at $\tau_{\mathrm{init}}$ and linearly decreased to a final value $\tau_{\mathrm{final}}$ over a predefined number of environment steps. After the annealing phase, $\tau$ remains fixed at $\tau_{\mathrm{final}}$ for the remainder of training. This annealing schedule is applied globally and synchronously across all agents.

Importantly, exploration in this setting is controlled at the population level rather than individually per agent. All agents share the same temperature parameter at any given time step, ensuring that the degree of stochasticity in action selection is homogeneous across the population. This design allows exploration strength to be treated as a single global control variable in the learning system.

To facilitate quantitative comparisons across annealing schedules, we summarize the
effective exploration regime by a scalar $B$, defined as the mean temperature over the
initial portion of training during which exploration is most active. Specifically,
\[
B = \frac{1}{K}\sum_{t=1}^{K}\tau(t), \qquad
K = \left\lfloor \frac{T_{\mathrm{anneal}}}{2}\right\rfloor,
\]
where $T_{\mathrm{anneal}}$ denotes the total number of temperature-annealing steps. This scalar provides a compact characterization of the overall exploration regime induced by a given annealing schedule, while abstracting away fine-grained temporal details of $\tau(t)$. Unless otherwise stated, all reported results are indexed by this exploration strength parameter.

\subsection{Simulation Protocol and Implementation Details}

We now describe the simulation protocol and implementation details used throughout the experiments, and summarize the associated learning procedure and hyperparameter settings. All results reported in this work are obtained under a fully synchronous multi-agent interaction protocol, with centralized learning through a shared replay buffer and a shared value network.

Unless otherwise stated, we set $L = 30$, resulting in $N = 900$ agents. Each experiment consists of a fixed number of discrete environment steps and is divided into two distinct phases: a training phase with annealed exploration, followed by a separate evaluation phase with learning disabled. Each run consists of a total of $T = 10^5$ environment steps, comprising a training phase of $T_{\mathrm{train}} = 95{,}000$ steps and a subsequent evaluation phase of $T_{\mathrm{eval}} = 5{,}000$ steps. During training, all agents observe their local states and select actions according to the softmax (Boltzmann) exploration policy defined in Eq.~\eqref{eq:softmax_policy}, with the exploration temperature $\tau$ annealed over time. Agent interactions generate experience tuples that are stored in a shared replay buffer and used to update the shared Q-network via temporal-difference learning. To improve credit assignment, $n$-step returns are employed whenever valid transition sequences are available, with a fallback to one-step targets otherwise. Target network parameters are synchronized with the online network at fixed intervals.

After the completion of the training phase, learning updates are disabled entirely. The system then enters the evaluation phase, during which the exploration temperature is fixed at $\tau_{\mathrm{eval}}$ and no gradient updates or replay sampling are performed. All reported performance metrics are computed exclusively over this evaluation period to ensure an unbiased assessment of the learned policy. To quantify collective behavior, we measure cooperation using an action-based order parameter defined as the fraction of cooperative actions executed by the population. Specifically, at each environment step $t$, we compute
\begin{equation}
c(t) = \frac{1}{N}\sum_{i=1}^{N} \mathbb{I}\!\left[a_i(t)=C\right],
\label{eq:coop_instant}
\end{equation}
where $\mathbb{I}[\cdot]$ denotes the indicator function and $a_i(t)\in\{C,D\}$ is the action selected by agent $i$ at time $t$. This definition directly captures realized behavior and avoids ambiguities associated with classifying agents under stochastic or mixed strategies.

The reported cooperation level is obtained by averaging $c(t)$ over the evaluation phase,
\begin{equation}
\bar{C} = \frac{1}{T_{\mathrm{eval}}}\sum_{t=1}^{T_{\mathrm{eval}}} c(t),
\label{eq:coop_avg}
\end{equation}
where $T_{\mathrm{eval}}=5{,}000$ environment steps correspond to the final evaluation period with fixed exploration temperature $\tau_{\mathrm{eval}}$ and learning disabled. This time-averaged quantity serves as our primary order parameter, characterizing steady-state cooperative behavior after transient dynamics have subsided.

All reported results are averaged over multiple independent simulation runs
with 30 different random seeds. Each run uses identical hyperparameters,
network architecture, and exploration schedules, differing only in the
initial random seed that governs network initialization, action sampling,
and environment stochasticity. Unless otherwise stated, each data point
shown in the figures corresponds to the mean cooperation level computed
over these independent seeds, ensuring that the reported patterns reflect
robust learning behavior rather than idiosyncratic stochastic effects.

Algorithm~\ref{alg:shared_dqn} provides a step-by-step description of the complete training and evaluation procedure, while Table~\ref{tab:hyperparams} summarizes the network architecture, optimization settings, and all hyperparameters used across experiments. Unless explicitly varied, these settings are held fixed throughout the study.

\begin{algorithm}[H]
\caption{Two-phase interaction protocol for shared-policy DQN.}
\label{alg:shared_dqn}
\begin{algorithmic}[1]
\For{$t = 1,2,\ldots,T$}
    \State Observe local states $\{s_i^t\}_{i=1}^N$.
    \If{$t \le T_{\mathrm{train}}$}
        \State Select actions $\{a_i^t\}$ using shared policy with annealed exploration $\tau(t)$.
    \Else
        \State Select actions $\{a_i^t\}$ using shared policy with fixed exploration $\tau_{\mathrm{eval}}$.
    \EndIf
    \State Execute actions and observe rewards $\{r_i^t\}$ and next states $\{s_i^{t+1}\}$.
    \If{$t \le T_{\mathrm{train}}$}
        \State Store transitions in shared replay buffer.
        \State Update shared Q-network from replayed samples.
        \State Periodically update target network and anneal $\tau(t)$.
    \EndIf
    \State Record evaluation metrics if $t > T_{\mathrm{train}}$.
\EndFor
\end{algorithmic}
\end{algorithm}

\begin{table}[H]
\centering
\caption{Learning hyperparameters and implementation details.}
\label{tab:hyperparams}
\begin{tabular}{l c}
\hline
\textbf{Parameter} & \textbf{Value} \\
\hline
Network architecture & 1 hidden layer, 96 units (ReLU activation) \\
Optimizer & AdamW \\
AdamW Learning rate & $1 \times 10^{-4}$ \\
AdamW Weight decay & $1 \times 10^{-4}$ \\
AdamW betas $(\beta_1, \beta_2)$ & $(0.9,\;0.999)$ \\
AdamW epsilon $\epsilon$ & $1 \times 10^{-8}$ \\
AdamW amsgrad & False \\
AdamW maximize & False \\
Discount factor $\gamma$ & 0.99 \\
Replay buffer size & 90{,}000 \\
Mini-batch size & 256 \\
Target network update & Every 2000 environment steps \\
$n$-step return & $n = 5$ (fallback to 1-step) \\
Loss function & Huber loss \\
Gradient clipping & $\ell_2$ norm $\leq 0.5$ \\
Exploration policy & Softmax (Boltzmann) \\
Annealing steps & 95{,}000 \\
Evaluation temperature $\tau_{\mathrm{eval}}$ & 0.10 \\
\hline
\end{tabular}
\end{table}

\section{Results}

We organize the results to progressively uncover why shared-policy Deep Q-Networks (DQN) fail in cooperative multi-agent environments under exploration. We use cooperation level as an indicator of learning stability and coordination quality. We systematically relate cooperation to internal representation dynamics, value landscapes, and environmental structure.

We establish five main findings. First, cooperation collapses systematically as exploration and payoff harshness increase (Figure~\ref{fig:heatmap_shared}). Second, this collapse follows empirically identifiable thresholds that differ between shared and grouped architectures (Figure~\ref{fig:boundary_shared_group}). Third, state augmentation that exposes learning progress can delay or partially reverse the collapse (Figure~\ref{fig:state_augmentation}). Fourth, cooperation collapse coincides with degradation in learned representations (Figure~\ref{fig:hidden_representation}) and contraction of action-value gaps (Figure~\ref{fig:q_gap}). Fifth, network topology modulates these effects: spatial structure buffers collapse while random topologies eliminate cooperation (Figure~\ref{fig:topology}).

Moving beyond behavioral outcomes, we analyze the internal representations learned by the shared network and show that cooperation collapse coincides with a degradation of feature separability between cooperative and defective behaviors (Figure~\ref{fig:hidden_representation}). At the value level, we further reveal a systematic contraction of action-value gaps as exploration increases, indicating convergence toward low-cooperation attractors rather than unstable or divergent learning dynamics (Figure~\ref{fig:q_gap}). Finally, we examine how interaction topology modulates these effects, showing that spatially structured environments provide implicit coordination scaffolding that buffers exploration-induced spillovers, whereas randomized topologies largely eliminate cooperative regimes (Figure~\ref{fig:topology}). Together, these results provide a coherent, multi-level characterization of a fundamental failure mode of shared DQN under non-stationary multi-agent interactions.

\subsection{Cooperation collapse across exploration and payoff parameters}

\begin{figure}[H]
   \centering
   \includegraphics[width=0.9\textwidth]{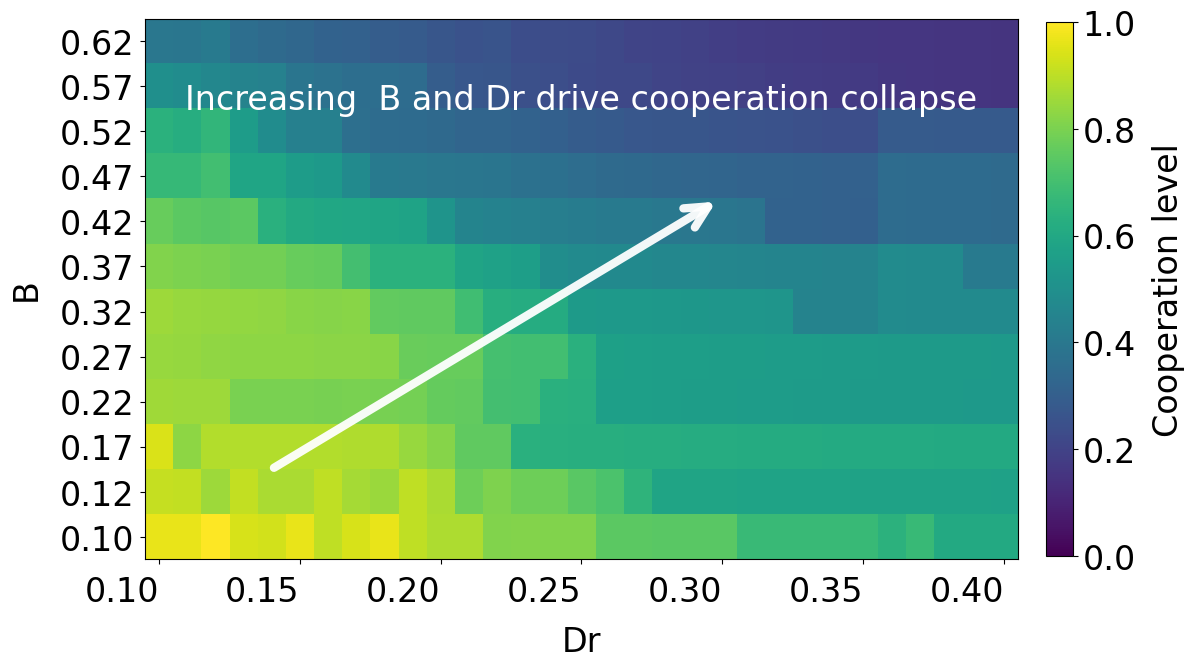}
\caption{Cooperation levels in shared-policy DQN as a function of exploration strength $B$ and payoff harshness $D_r$. Each cell is the mean over 30 random seeds during evaluation. Cooperation degrades systematically as either parameter increases, forming an extended low-cooperation region in the upper-right. The smooth boundaries indicate structured collapse rather than sporadic failures.}
   \label{fig:heatmap_shared}
\end{figure}

Figure~\ref{fig:heatmap_shared} shows cooperation levels across two dimensions: exploration strength $B$ and payoff harshness $D_r$. Each cell is the mean over 30 random seeds during the evaluation phase. This representation allows us to directly visualize how cooperative behavior evolves as exploration and incentive asymmetry jointly increase. Cooperation degrades monotonically as either $B$ or $D_r$ increases. Low-cooperation outcomes form an extended region in the upper-right of the parameter space. The transition is continuous, not abrupt. This indicates structured collapse rather than sporadic failures from random initialization.

The collapse region expands smoothly with increasing exploration. At low $B$, cooperation remains high even when payoffs favor defection. Beyond a critical $B$, cooperation drops rapidly as $D_r$ increases. Neither factor alone determines the failure. The collapse depends on their interaction. 

\subsection{Empirical collapse thresholds and the role of representation sharing}

We summarize the collapse using an empirical threshold $D_r^*(B)$ for each exploration level. This is the largest $D_r$ value (within our tested range $[0.10, 0.40]$) where cooperation remains above a criterion level. We average over 30 random seeds to ensure robust thresholds.

\begin{figure}[H]
   \centering
   \includegraphics[width=1\textwidth]{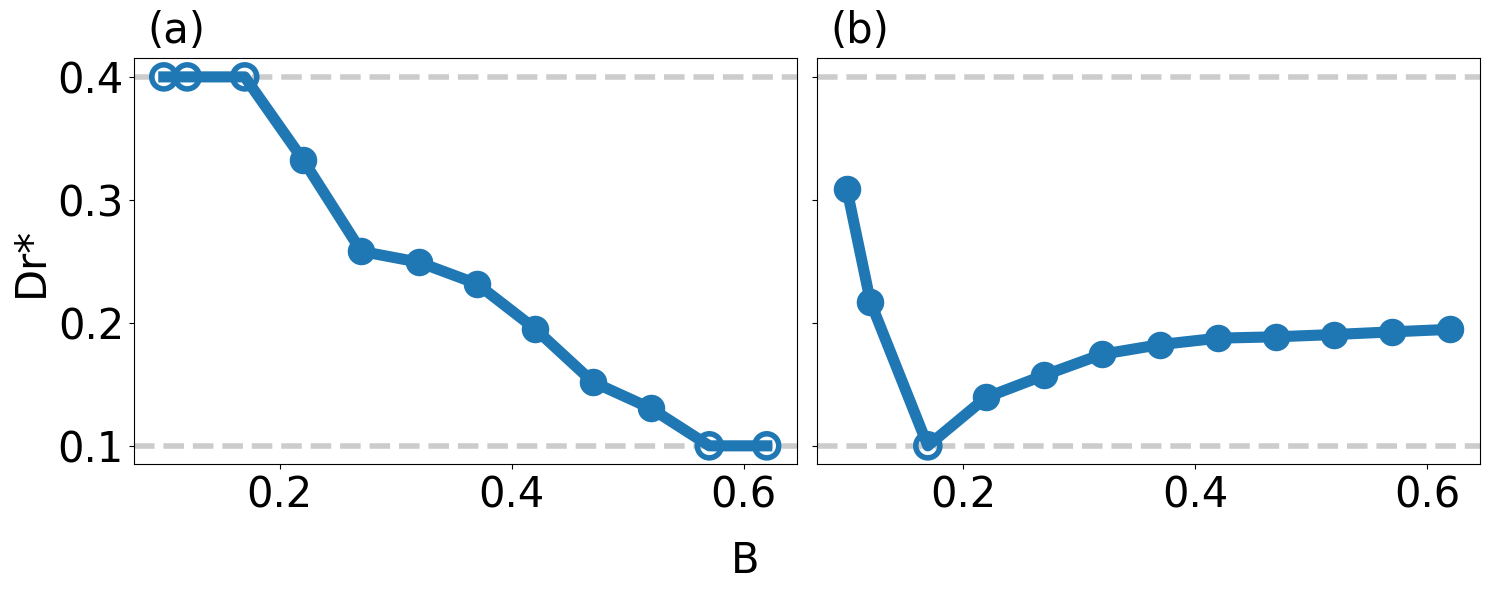}
\caption{Collapse thresholds $D_r^*(B)$ as a function of exploration strength $B$ (mean over 30 seeds). (a) Shared DQN: all agents use one network. Threshold decreases monotonically with $B$ (identified at cooperation = 0.55). (b) Grouped DQN: 10 groups of 90 agents, each with independent network. Threshold shows non-monotonic dependence with partial recovery at high $B$ (identified at cooperation = 0.15). Different thresholds reflect different baseline cooperation levels. Dashed lines mark tested $D_r$ range $[0.10, 0.40]$. Hollow markers indicate threshold outside tested range.}
   \label{fig:boundary_shared_group}
\end{figure}

Figure~\ref{fig:boundary_shared_group} reports the resulting collapse thresholds for two learning architectures. Figure~\ref{fig:boundary_shared_group}(a) corresponds to the fully shared-policy DQN, in which all agents update a single common value network. Figure~\ref{fig:boundary_shared_group}(b) shows a grouped baseline, where agents are randomly partitioned into ten disjoint groups, each maintaining an independent shared value function. This grouped design limits the scope of representation sharing while preserving within-group coordination, providing an intermediate comparison between fully shared and fully independent learning.

For the shared-policy DQN, the extracted threshold $D_r^*(B)$ decreases monotonically as exploration increases. Higher exploration progressively reduces the range of payoff conditions under which cooperation can be sustained, yielding a smooth and systematic shift of the collapse threshold across $B$. Importantly, this trend emerges consistently across the explored parameter space, rather than as isolated or irregular breakdowns. Grouped DQN shows a different pattern. At low $B$, the threshold decreases like shared DQN. However, at high $B$, it stabilizes and partially recovers. Limiting representation sharing to smaller groups reduces the impact of exploration. The two architectures use different cooperation thresholds to define collapse: 0.55 for shared DQN and 0.15 for grouped DQN. This is necessary because grouped DQN achieves much lower baseline cooperation (see Figure~\ref{fig:boundary_shared_group}). Both thresholds are chosen so the crossing occurs within our tested $D_r$ range $[0.10, 0.40]$. These different thresholds do not affect the qualitative comparison, as we are comparing the trends (how thresholds change with $B$), not the absolute values.

Taken together, these results indicate that cooperation collapse in shared DQN admits a structured empirical characterization via collapse thresholds, and that the monotonic deterioration observed in the fully shared architecture is strongly attenuated when representation sharing is restricted. This contrast highlights the central role of representation aggregation in shaping how exploration-induced non-stationarity manifests at the system level.

\subsection{State augmentation delays cooperation collapse}

Does cooperation collapse because agents cannot observe the learning dynamics? We test this by augmenting the state with signals about exploration and training progress. Under this baseline, cooperation declines steadily as $B$ increases.

\begin{figure}[H]
   \centering
   \includegraphics[width=0.9\textwidth]{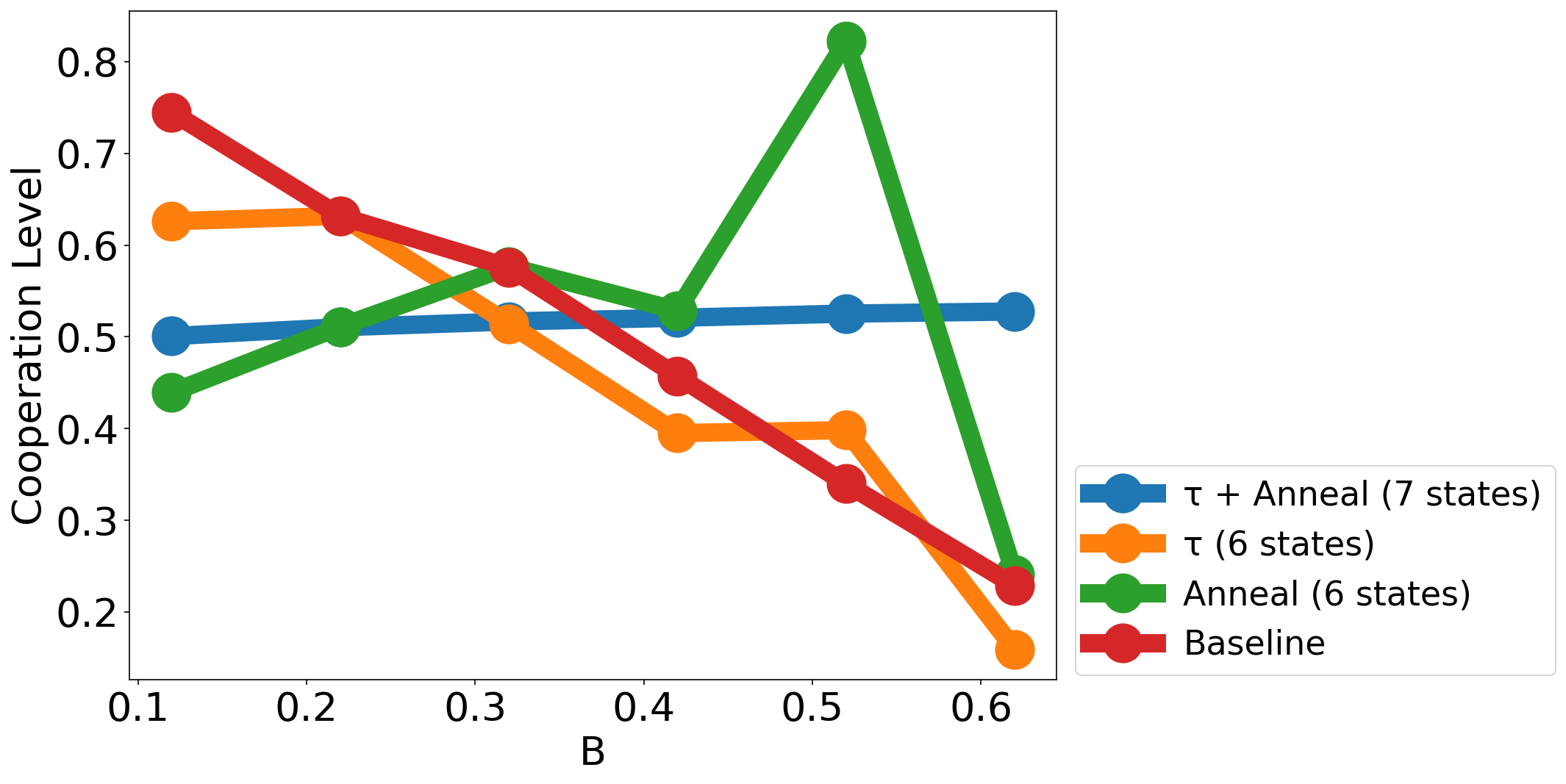}
   \caption{Effect of state augmentation on cooperation outcomes in shared-policy DQN under varying exploration strength $B$. All curves show mean cooperation levels averaged over 30 independent random seeds. The baseline curve corresponds to the original shared DQN, whose state encodes only local interaction information, consisting of each agent’s own recent action together with the recent actions of its neighbors. We compare this baseline against augmented variants that append additional scalar signals to the original state vector: a temporal exploration signal $\tau$, a coarse training-progress (annealing) indicator, and their joint inclusion. All configurations employ identical learning algorithms, network architectures, reward structures, and interaction topologies; only the information available to the shared value function differs. Joint augmentation yields the most robust behavior relative to the baseline, exhibiting a stable and weakly increasing cooperation trend as exploration strength $B$ increases, while substantially reducing sensitivity to exploration-induced degradation. In contrast, single-signal augmentation produces more limited and regime-dependent effects.
}
   \label{fig:state_augmentation}
\end{figure}

The baseline shared DQN uses a minimal state: the last actions of the agent's four neighbors plus its own last action. This is a 5-dimensional binary vector with no information about time, training progress, or exploration level. Under this baseline, cooperation declines steadily as $B$ increases (red curve in Figure~\ref{fig:state_augmentation}). Relative to the baseline, different forms of state augmentation exhibit distinct effects on cooperation under exploration. The baseline shared DQN shows a monotonic decline in cooperation as the exploration strength $B$ increases, reflecting a progressive breakdown of coordination. In contrast, augmenting the state with a temporal exploration signal $\tau$ yields modest improvements at low to intermediate exploration levels, but cooperation deteriorates rapidly as $B$ increases further, eventually falling below the baseline at high exploration. This indicates that temporal information alone is insufficient to robustly stabilize shared learning.

Augmenting the state with a training-progress signal produces a qualitatively different pattern. Cooperation is substantially enhanced within a narrow range of intermediate exploration levels, but this improvement is highly regime-dependent and unstable, with cooperation sharply declining again at higher $B$. By comparison, the joint augmentation that incorporates both temporal and training-progress signals provides the most robust stabilization. Relative to the baseline, cooperation under joint augmentation exhibits a stable and slightly increasing trend as the exploration strength $B$ increases, while remaining markedly less sensitive to exploration-induced degradation. This behavior indicates that jointly encoding multiple forms of latent temporal structure is critical for stabilizing shared representations under exploration.

Taken together, these results provide causal evidence that the cooperation collapse observed in the baseline shared DQN is tightly linked to partial observability of non-stationary learning dynamics. By explicitly exposing latent temporal signals to the shared representation, state augmentation significantly enhances robustness under exploration, demonstrating that the failure is not solely driven by payoff incentives or algorithmic instability, but by representational limitations induced by insufficient state information.

\subsection{Representation collapse in the shared network under exploration}

\begin{figure}[H]
   \centering
   \includegraphics[width=1\textwidth]{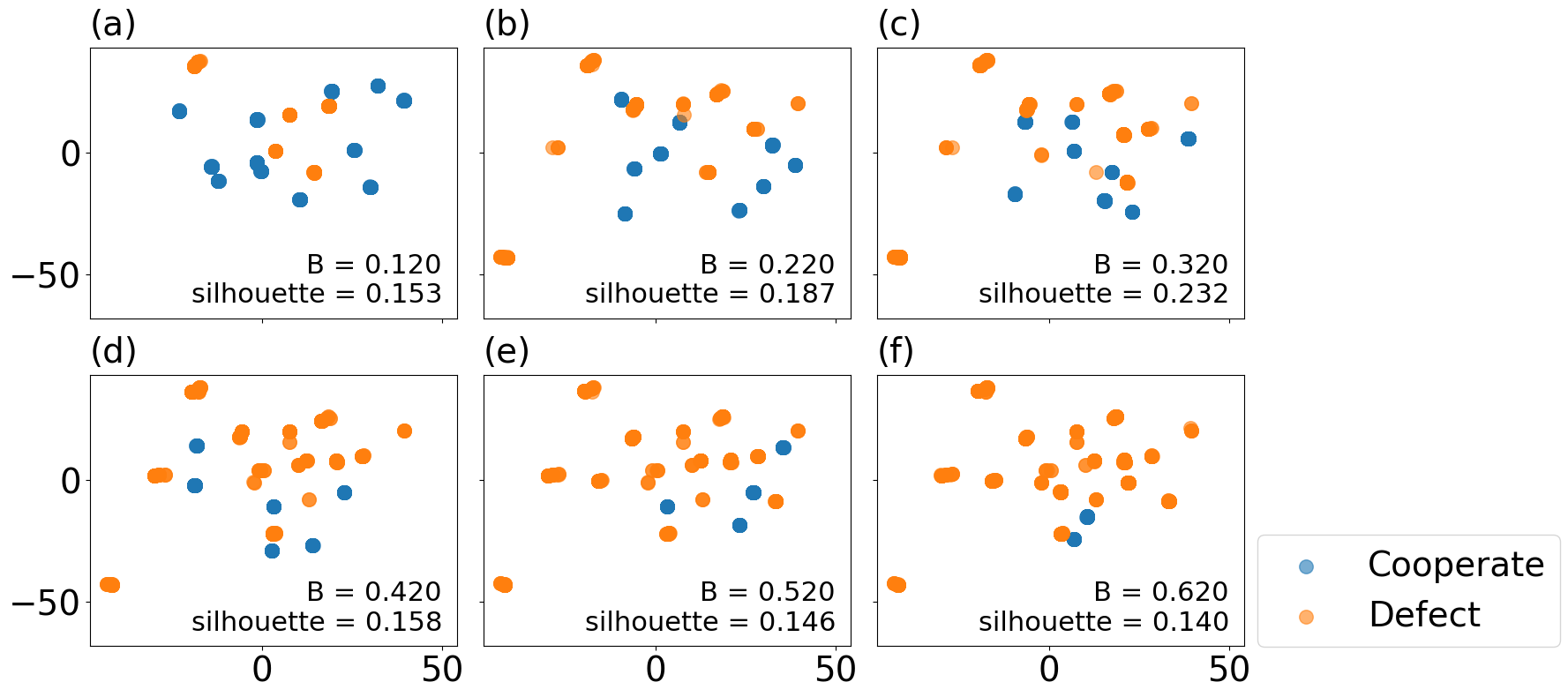}
   \caption{Visualization of hidden-layer representations learned by the shared-policy DQN under increasing exploration strength $B$. Each panel shows a two-dimensional UMAP projection of hidden activations corresponding to agents’ states, with points colored by the selected action (cooperate or defect). At low exploration, cooperative and defective actions occupy partially distinct regions in the projected latent space. At intermediate $B$, the representations exhibit increased geometric clusterability, while under strong exploration the latent space becomes increasingly diffuse, with cooperative representations shrinking and losing stability. Silhouette scores, computed from unsupervised two-cluster partitioning of the hidden representations, quantify this non-monotonic evolution in cluster structure. Importantly, silhouette scores reflect geometric clusterability rather than action-discriminative separability, and their transient increase at intermediate $B$ does not correspond to improved support for cooperative decision-making.}
   \label{fig:hidden_representation}
\end{figure}

Cooperation collapse can be delayed by reducing representation sharing or by state augmentation. We now examine the internal representations learned by the shared network. Do changes in cooperation correspond to changes in learned features? To this end, we analyze the hidden-layer activations of the shared DQN associated with cooperative and defective actions across different levels of exploration strength $B$. Figure~\ref{fig:hidden_representation} visualizes these representations using UMAP, a nonlinear dimensionality reduction technique employed here solely for qualitative inspection of representational geometry. Points are colored according to the agent’s selected action. To quantitatively characterize the organization of the hidden representations, we compute silhouette scores based on an unsupervised two-cluster partitioning of the hidden-layer activations. Specifically, for each exploration level $B$, the hidden representations are partitioned into two clusters using $k$-means clustering with $k=2$, without using action labels. For each representation vector $i$, the silhouette coefficient is defined as
\[
s(i) = \frac{b(i) - a(i)}{\max\{a(i), b(i)\}},
\]
where $a(i)$ denotes the average distance between $i$ and other points within the same cluster, and $b(i)$ denotes the minimum average distance between $i$ and points in the nearest neighboring cluster. The silhouette score reported for each $B$ is obtained by averaging $s(i)$ over all sampled representations.

Silhouette scores show non-monotonic evolution with $B$. At intermediate $B$, scores increase as representations cluster into two groups. However, this does not mean better action discrimination. It reflects convergence toward a defection-dominated regime where most states map to similar features. Under strong exploration, this structure deteriorates. The latent space becomes increasingly diffuse, with cooperative representations shrinking into sparse and unstable regions while defective representations dominate the manifold. Correspondingly, silhouette scores decline at high $B$, indicating a loss of coherent cluster structure in the hidden space. Importantly, this breakdown occurs despite numerically stable training and apparent convergence, ruling out optimization failure or divergence as its cause.

The non-monotonic behavior of silhouette scores contrasts with the monotonic decline in cooperation rates observed at the behavioral level. This discrepancy highlights a critical distinction between clusterability of hidden representations and action-discriminative separability. While intermediate exploration may transiently increase representational consistency around a dominant defection-oriented regime, it does not enhance the network’s capacity to support cooperative decision-making.

Taken together with the intervention results in Figure~\ref{fig:state_augmentation}, these findings indicate that the primary vulnerability of shared DQN lies at the level of representation learning. When exploration-induced non-stationarity is aggregated across agents without explicit temporal encoding or structural decoupling, the shared network is unable to maintain stable, action-aligned internal representations, ultimately undermining sustained cooperation.

\subsection{Value landscape contraction and convergence to low-cooperation attractors}

The preceding analyses demonstrate that cooperation collapse in shared DQN is
accompanied by a degradation of internal representations. We now investigate
whether this collapse is associated with unstable or divergent learning
dynamics, or instead reflects convergence toward a qualitatively different and
compressed value landscape. All value statistics reported in this section are
computed by averaging over multiple independent simulation runs with 30 different
random seeds, ensuring that the observed contraction of the value landscape
reflects robust convergence behavior rather than seed-specific artifacts.

\begin{figure}[H]
   \centering
   \includegraphics[width=1\textwidth]{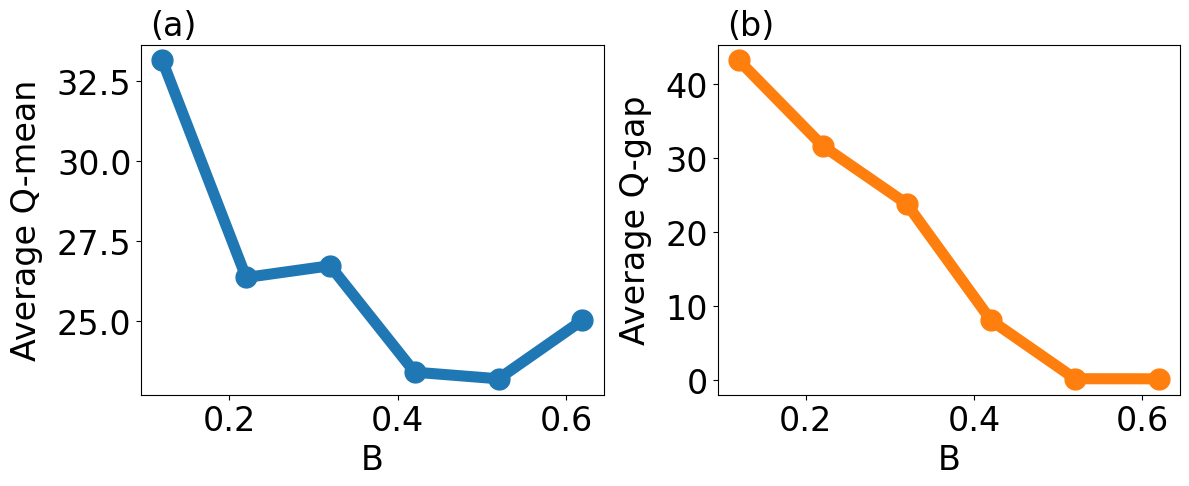}
   \caption{Average action-value statistics of the shared-policy DQN as functions of the exploration strength $B$, evaluated during a held-out evaluation phase with learning disabled, with the defection loss fixed at $D_r = 0.25$. All statistics are averaged over 30 independent random seeds. (a) Average action-value magnitude (Q-mean), computed as the mean absolute action value averaged over actions, agents, and encountered states. (b) Average action-value gap (Q-gap), defined as the mean absolute difference between the estimated Q-values of cooperative and defective actions. As exploration increases, the Q-gap decreases monotonically toward zero, while the Q-mean remains finite and bounded, indicating a contraction of relative value differences rather than unstable or divergent learning dynamics.}
   \label{fig:q_gap}
\end{figure}

Figure~\ref{fig:q_gap} reports two complementary statistics as functions of the exploration strength $B$, with the defection loss fixed at $D_r = 0.25$: (a) the average action-value magnitude (Q-mean), defined as the mean absolute action value averaged over actions, agents, and encountered states, and (b) the average action-value gap (Q-gap) between cooperative and defective actions. The Q-gap is defined as the mean absolute difference between the estimated Q-values of the two actions. Both quantities are computed during a held-out evaluation phase with learning disabled, ensuring that the reported statistics reflect the geometry of the converged value function rather than transient training dynamics.

As $B$ increases, the Q-gap exhibits a clear and monotonic decline, approaching zero at high exploration levels. This trend indicates a systematic erosion of value discrimination between cooperative and defective actions. In contrast, the Q-mean does not decrease monotonically: after an initial reduction, it stabilizes within a bounded range and exhibits mild non-monotonic variations. Crucially, the Q-mean remains finite and well-behaved across all exploration regimes, showing no signs of divergence or oscillatory instability.

Together, these observations rule out training instability or numerical divergence as the primary drivers of cooperation collapse. Instead, the dominant effect of increasing exploration is a contraction of the relative value landscape: while absolute action-value magnitudes persist, the differential advantage between actions progressively vanishes. As a result, the shared DQN converges reliably to policies that are stable but weakly discriminative, in which cooperative actions no longer enjoy a robust value advantage.

This convergence corresponds to low-cooperation attractors characterized by compressed action-value differences rather than erratic learning dynamics. In this regime, exploration reshapes the learned value geometry without destabilizing optimization, leading to indifference between strategic alternatives.

Taken together with the representation-level analysis in Figure~\ref{fig:hidden_representation}, these results suggest that cooperation collapse in shared DQN arises from a coordinated degradation of value discrimination and latent structure, rather than from failures of convergence or numerical stability.

\subsection{Topology robustness of exploration-induced cooperation collapse}

We next assess the robustness of the exploration-induced cooperation collapse
in the shared-policy DQN across distinct interaction topologies, including grid,
modular 4-regular, random 4-regular, and small-world networks. For all considered
topologies, we fix the average node degree to 4, such that differences in
cooperation dynamics can be attributed solely to structural organization rather
than connectivity density. All cooperation curves reported in this section are
averaged over multiple independent simulation runs with 30 different random seeds,
ensuring that the observed cross-topology differences reflect robust structural
effects rather than seed-specific fluctuations.

\begin{figure}[H]
   \centering
   \includegraphics[width=0.85\textwidth]{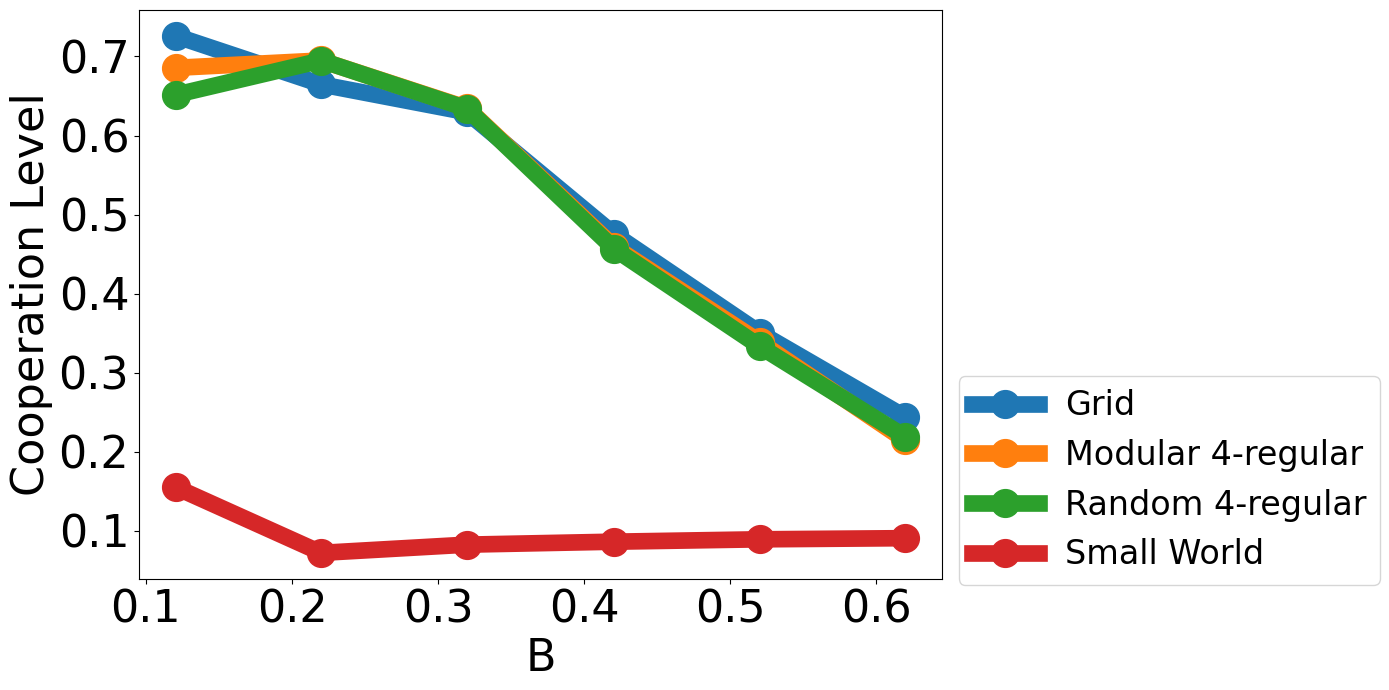}
   \caption{Average cooperation level as a function of the exploration strength $B$ under the shared-policy DQN across different interaction topologies, including grid, modular 4-regular, random 4-regular, and small-world networks. All curves are averaged over 30 independent random seeds. All networks are constructed with an average degree of 4 and the same total number of agents, ensuring a controlled comparison across interaction structures. All results are evaluated at a fixed temptation parameter $D_r = 0.25$ and computed over the evaluation phase after training convergence. All topologies are instantiated with the same total number of agents as the grid network to isolate the effect of interaction structure from population size. While the grid, modular, and random topologies exhibit nearly overlapping cooperation curves with a smooth decline as exploration increases, the small-world topology sustains uniformly low cooperation across the entire range of $B$, indicating heightened vulnerability of shared DQN to exploration under globally mixed interaction structures.
}
   \label{fig:topology}
\end{figure}

Figure~\ref{fig:topology} reports the average cooperation level as a function of the exploration strength $B$, evaluated at a fixed temptation parameter $D_r = 0.25$ and computed over the evaluation phase after training convergence. To ensure a fair comparison across topologies, all networks are constructed to contain the same total number of agents as the grid baseline, differing only in their connectivity patterns. Across the grid, modular 4-regular, and random 4-regular topologies, we observe a consistent qualitative pattern: cooperation remains relatively high at low exploration levels ($B \leq 0.2$), followed by a pronounced and monotonic decline as $B$ increases. Remarkably, the cooperation curves for these three locally homogeneous topologies nearly overlap throughout the entire range of $B$, indicating that moderate variations in local connectivity—whether regular, modular, or random—do not substantially alter the collapse dynamics of shared DQN.

In contrast, the small-world topology exhibits a markedly different behavior. Cooperation levels remain uniformly low across all exploration strengths, with only minor variation as $B$ increases. Unlike the other topologies, cooperative behavior fails to emerge even at low exploration, suggesting that the presence of long-range connections disrupts the formation of stable local coordination under a globally shared value function.

Taken together, Figure~\ref{fig:topology} demonstrates that the collapse of cooperation in shared-policy DQN is largely insensitive to local structural variations, but can be significantly exacerbated by topologies that promote rapid global mixing. These results support the interpretation that the observed collapse reflects intrinsic limitations of shared representation learning under exploration, rather than artifacts of specific network constructions. This complements our earlier evidence that interaction-graph organization and adaptation mechanisms critically shape cooperative regime formation in learning-driven social dilemmas \citep{weng2025q}.

\section{Discussion}

\subsection{What fails and what does not}

Cooperation collapse in shared-policy DQN does not result from unstable learning or numerical divergence. Learning remains well-behaved across all tested conditions. Value estimates stay bounded, training converges reliably, and evaluation metrics are stable. The failure is not whether the system converges, but what it converges to. Shared DQN reliably finds stable solutions that fail to sustain cooperation. This distinction matters. Performance degradation is often blamed on excessive noise, poor tuning, or optimization issues. Our results rule out all three. The collapse persists across varied learning rates, loss functions, optimizers, and buffer sizes (see Appendix). Action-value gaps contract smoothly and monotonically, not erratically.

The failure is not simply about incentives. Even at fixed intermediate payoffs, cooperation collapses as exploration increases. Changing incentives alone cannot explain the pattern. This rules out a simple equilibrium shift. The collapse emerges from the learning dynamics themselves. This echoes early findings that independent learners struggle without coordination \cite{tan1993multi}.

We reframe cooperation collapse as a structural failure mode. It arises from how shared representations handle non-stationary multi-agent learning under exploration. The tension between independent learning and coordination has been extensively documented in cooperative Markov games \cite{matignon2012independent}, where agents using standard single-agent algorithms often fail to converge to globally optimal solutions.

\subsection{Shared representations as a structural bottleneck}

Shared-policy learning has a structural tension. All agents contribute to and use one value function, despite experiencing different local contexts and learning at different rates. This improves scalability and sample efficiency. However, it aggregates non-stationarity across agents instead of canceling it out. This aggregation of learning signals represents a fundamental challenge in multi-agent deep reinforcement learning, where the environment becomes non-stationary from each agent's perspective \cite{hernandez2019survey}. Each agent's exploration perturbs its neighbors' environments. These perturbations flow through the shared replay buffer into gradient updates. The shared network receives conflicting signals from agents at different learning stages. As exploration increases, these conflicts intensify and cannot be resolved within a single representation.

Our representation-level analyses reveal the consequences of this aggregation. Cooperation collapse coincides with degradation of action-aligned latent structure and a systematic contraction of action-value differences, indicating convergence toward stable but weakly discriminative value landscapes. This convergence does not indicate optimization failure. Instead, it reflects representational interference: the shared network must accommodate incompatible signals from different agents within fixed capacity~\cite{atkinson2021pseudorehearsal}.

The failure is a representation-level bottleneck. Well-defined collapse thresholds emerge in parameter space, similar to phase transitions in coupled dynamical systems \cite{lee2025phase}. This suggests the failure is an architectural feature, not a tuning artifact. Collapse occurs when one representation cannot maintain discriminative features across agents in different interaction contexts. 

\subsection{Partial observability amplifies the failure}

Partial observability worsens the representation bottleneck. Standard states encode only recent actions, assuming the environment is stationary. This assumption fails under concurrent learning.

The same local state can mean different things at different training stages. Early in training, four cooperating neighbors might indicate stable cooperation. Late in training, the same observation might indicate temporary noise in a defection regime. Without temporal context, the network cannot distinguish these cases. This intensifies representational interference. State augmentation provides causal evidence. Adding exploration level and training progress to the state improves observability without changing the algorithm. This delays cooperation collapse across all exploration levels and partially reverses it in some regimes. The intervention works, proving that state information matters.

However, state augmentation is a diagnostic, not a solution. It reduces conflict but does not eliminate coupling. All agents still share one network. Partial observability amplifies the structural limitation but is not the only cause.

\subsection{Why mitigation helps but cannot eliminate failure}

Mitigation strategies alleviate but do not resolve the failure. Grouped learning limits parameter sharing. State augmentation improves observability. Both extend the range where cooperation survives, but neither eliminates the fundamental constraint.

The structural constraint remains. Even with augmented states, all agents use one value function. Gradients from different agents still interfere. Experience replay adds complications: stored transitions reflect outdated policies of other agents \cite{foerster2017stabilising}. Grouped architectures reduce coupling but do not eliminate it. They trade global sharing for smaller-scale sharing. At high exploration, collapse persists despite mitigation. The shared representation eventually exceeds its capacity. No amount of state augmentation or partial decoupling can overcome this limit.

These findings clarify why cooperation collapse should be viewed as a fundamental limitation of shared-policy learning under strong non-stationarity, rather than as a deficiency addressable through incremental tuning or auxiliary inputs. Mitigation strategies expand the robustness margin, but cannot remove the representational bottleneck imposed by global parameter sharing.

\subsection{Implications for scalable multi-agent reinforcement learning design}

Our findings have four implications for MARL system design. First, parameter sharing is not free. It improves sample efficiency but aggregates non-stationarity. Designers must explicitly manage the trade-off between scalability and representational coherence.

Second, robustness requires observing learning dynamics, not just environment state. Standard Markovian states assume stationarity, which fails under concurrent learning. Conditioning policies on training phase or exploration level can improve stability.

Third, representation modularity is critical. The failure arises from forcing one representation to handle all agents and learning phases. Modular, hierarchical, or partially decoupled architectures may better localize non-stationarity.

Finally, convergence diagnostics in MARL should extend beyond reward trajectories. Our results show that stable convergence can coexist with qualitative degradation in value geometry and action discrimination. Monitoring representation structure and relative value landscapes is therefore essential for assessing robustness in shared-policy systems. 

\subsection{Scope, limitations, and applicability of the diagnosis}

We studied shared-policy DQN as a representative value-based MARL method. Many findings likely generalize to other architectures with full parameter sharing. However, methods with centralized critics, opponent modeling, or modular representations may avoid these failures.

The dynamic Prisoner's Dilemma is used as a diagnostic environment rather than a comprehensive benchmark. Its simplicity enables controlled manipulation of exploration, incentives, and topology, allowing isolation of representation-level mechanisms. While quantitative outcomes may vary across tasks, the qualitative tensions identified---between shared representations, non-stationarity, and exploration---are not specific to this game.

Importantly, the cooperation collapse documented in this work does not reflect numerical instability or failure to converge. In all regimes, shared DQN converges reliably to well-defined value functions. The limitation identified here is therefore qualitative, concerning the nature of the learned solution rather than the existence of one.

Our results reveal an underappreciated risk: shared representations struggle with non-stationary multi-agent learning. Understanding this failure mode can guide the design of more robust MARL systems.

\section*{Data availability}

The source code for all simulations and analyses is publicly available at the author's GitHub. Raw simulation data, network parameters, and reproducibility scripts are included in the repository. All results can be reproduced using the provided code and documented random seeds.

\bibliographystyle{unsrt}
\bibliography{ref}
\clearpage

\appendix
\renewcommand{\thefigure}{A.\arabic{figure}}
\setcounter{figure}{0}
\renewcommand{\thetable}{A.\arabic{table}}
\setcounter{table}{0}
\renewcommand{\theequation}{A.\arabic{equation}}
\setcounter{equation}{0}
\section{Supplementary Material}

\subsection*{Appendix A.1. Sensitivity to population size}

\begin{figure}[H]
    \centering
    \includegraphics[width=1\textwidth]{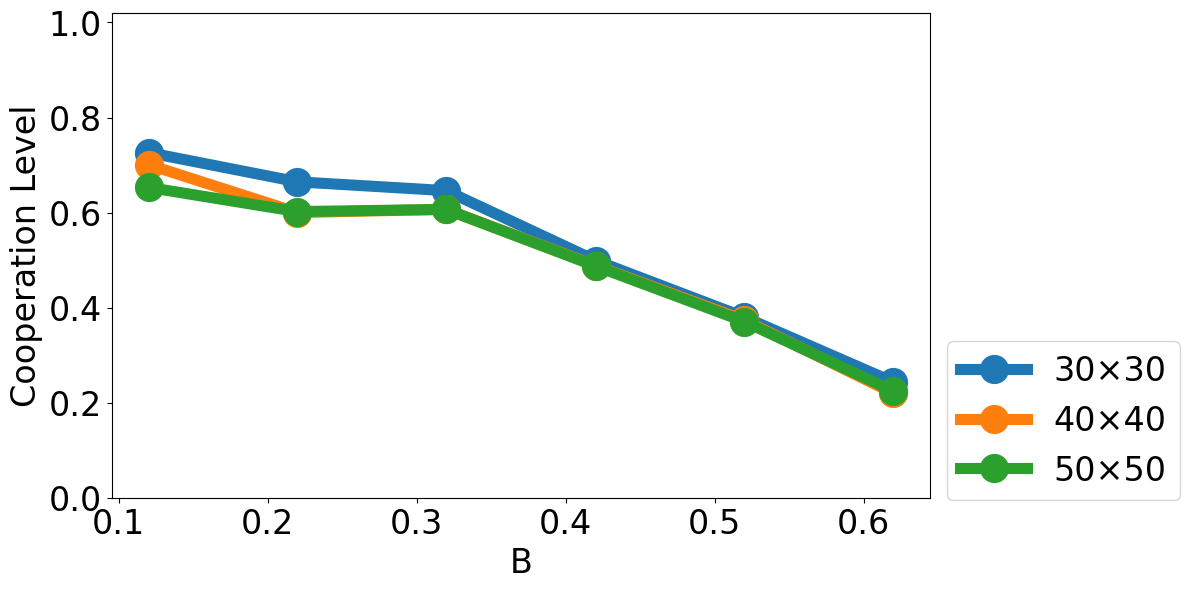}
    \caption{Sensitivity of cooperation outcomes to population size under shared-policy DQN. Average cooperation level during the evaluation phase is shown as a function of the exploration strength $B$ for three grid sizes: $30\times30$, $40\times40$, and $50\times50$. All results are computed after training convergence with learning disabled, at a fixed payoff parameter $D_r = 0.25$, using identical learning hyperparameters and exploration schedules across system sizes. Across all populations, cooperation decreases monotonically with increasing $B$, and the curves exhibit closely aligned collapse trajectories. Minor quantitative differences appear only at low exploration levels, while the onset and progression of cooperation collapse remain consistent at moderate and high $B$. This indicates that the exploration-induced cooperation collapse observed in shared-policy DQN is robust to population size and not driven by finite-size effects.
}
    \label{fig:size}  
\end{figure}

Figure~\ref{fig:size} examines the sensitivity of the shared-policy DQN to the population size by comparing average cooperation levels across three grid sizes, $30\times30$, $40\times40$, and $50\times50$, as a function of the exploration strength $B$. All results are evaluated during the fixed evaluation phase with learning disabled, using identical payoff parameters ($D_r = 0.25$) and learning hyperparameters across system sizes.

Across all three population scales, we observe a consistent qualitative pattern: cooperation remains relatively high at low exploration levels and declines monotonically as $B$ increases. The location and shape of this decline are remarkably stable across grid sizes, with only minor quantitative differences at small $B$. In particular, larger populations exhibit slightly lower cooperation in the low-exploration regime, reflecting increased heterogeneity in local interaction contexts, but the overall collapse trajectory remains unchanged.

Importantly, the onset and progression of cooperation collapse as exploration increases do not depend sensitively on the system size. The near-overlap of the curves at moderate and high $B$ indicates that the exploration-induced degradation observed in the main results is not a finite-size artifact, nor driven by small-population effects. Instead, it reflects an intrinsic property of shared-policy DQN under non-stationary multi-agent learning.

These results confirm that the structured cooperation collapse reported in the main text persists across different population scales, supporting the generality and scalability of the observed failure mode. Figure~\ref{fig:size} therefore serves as a robustness check, demonstrating that the conclusions drawn from the default $30\times30$ setting extend to substantially larger systems.

\subsection*{Appendix A.2. Sensitivity to learning hyperparameters}

\begin{figure}[H]
    \centering
    \includegraphics[width=1\textwidth]{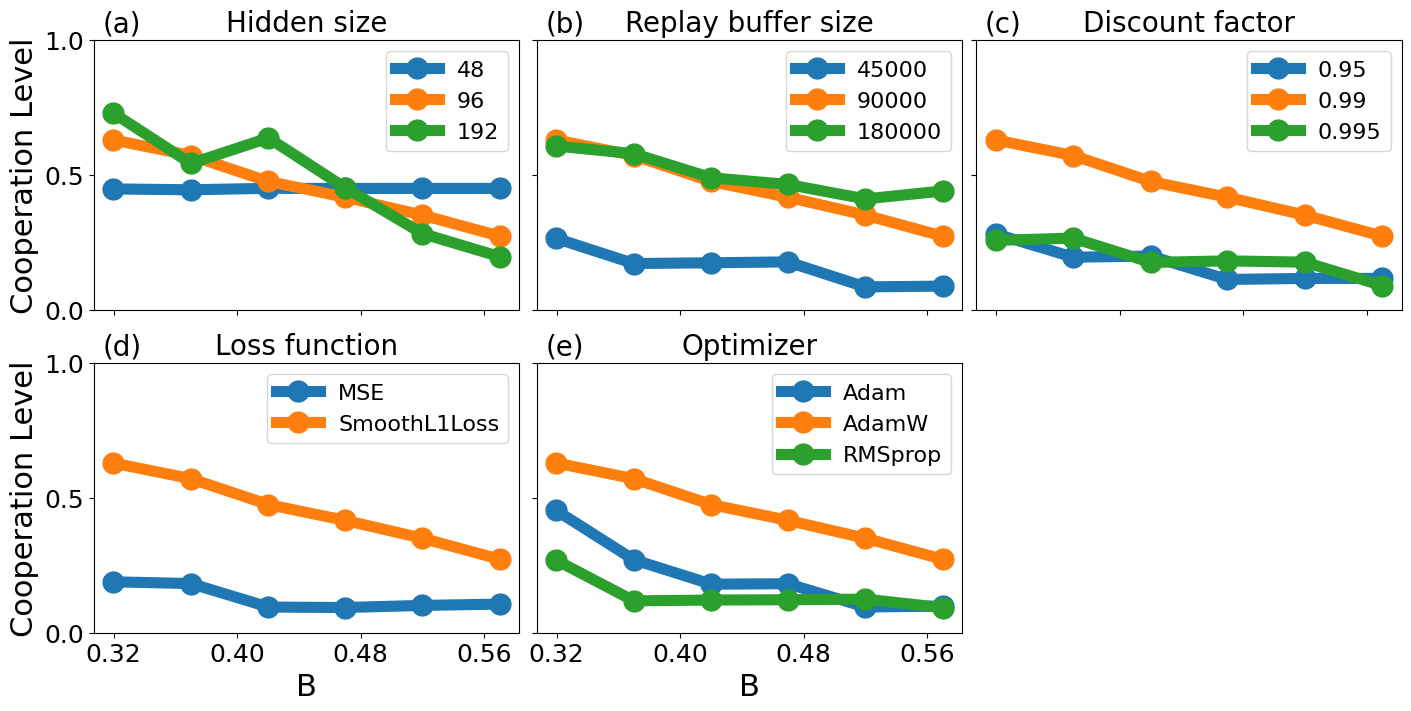}
    \caption{Sensitivity of cooperation outcomes in shared-policy DQN to learning hyperparameters under fixed payoff conditions. The average cooperation level during the evaluation phase is shown as a function of the exploration strength $B$ for variations in (a) hidden layer size, (b) replay buffer capacity, (c) discount factor $\gamma$, (d) loss function, and (e) optimizer. All results are computed after training convergence with learning disabled, at a fixed payoff parameter $D_r = 0.25$, using identical interaction topology, exploration schedules, and simulation protocol unless explicitly stated otherwise. Across all hyperparameter settings, cooperation decreases monotonically as $B$ increases. While different architectural and optimization choices shift the absolute cooperation level at low exploration, no configuration alters the qualitative dependence on $B$ or restores cooperation to the high-cooperation regime observed under weak exploration. 
}
    \label{fig:general sa}  
\end{figure}

Figure~\ref{fig:general sa} presents a systematic sensitivity analysis of the shared-policy DQN with respect to key learning hyperparameters, including (a) hidden layer size, (b) replay buffer capacity, (c) discount factor $\gamma$, (d) loss function, and (e) optimizer. For all panels, the average cooperation level is evaluated during a fixed evaluation phase with learning disabled, at a fixed payoff parameter $D_r = 0.25$. Unless otherwise stated, all remaining components of the learning algorithm, interaction topology, and simulation protocol are identical to those used in the main experiments.

Across all hyperparameter variations, a consistent qualitative pattern emerges: cooperation decreases monotonically as the effective exploration strength $B$ increases. While different architectural and optimization choices lead to noticeable quantitative shifts in cooperation levels at low exploration, the ordering and collapse trajectory with respect to $B$ remain unchanged. After averaging across 30 random seeds, no hyperparameter configuration alters the monotonic dependence on $B$ or restores cooperation to the high-cooperation regime observed under weak exploration.

Panel (a) shows that increasing the hidden layer size primarily affects the absolute cooperation level at low $B$, with larger networks exhibiting higher cooperation in low-exploration regimes. However, all network sizes converge toward similarly low cooperation levels as exploration increases, indicating that increased representational capacity does not prevent the exploration-induced collapse. Panel (b) demonstrates that replay buffer capacity shifts the cooperation scale but does not qualitatively modify the dependence on $B$: larger buffers consistently yield higher cooperation at low exploration, yet still exhibit a monotonic decline as $B$ increases.

Panel (c) indicates that varying the discount factor $\gamma$ shifts the overall cooperation level under weak exploration. However, the qualitative sensitivity to exploration remains identical across discount factors, and all settings converge to low-cooperation regimes at sufficiently large $B$. This suggests that differences in temporal credit assignment do not mitigate the collapse induced by strong exploration.

Panels (d) and (e) further confirm that neither the choice of loss function nor optimizer fundamentally alters the observed dynamics. SmoothL1 loss and AdamW consistently yield higher cooperation levels than MSE and Adam under low exploration, while RMSprop produces uniformly lower cooperation across all $B$. Importantly, all optimizers exhibit a monotonic decrease in cooperation as exploration increases, reinforcing that the observed collapse is not attributable to a specific optimization method.

Taken together, Figure~\ref{fig:general sa} demonstrates that the exploration-induced cooperation collapse in the shared-policy DQN is robust across a broad range of commonly used learning hyperparameters. These results confirm that the failure mode identified in the main text is not an artifact of architectural capacity, replay memory design, discounting, or optimization choices, but reflects an intrinsic limitation of shared representation learning under strongly non-stationary multi-agent interactions.

\subsection*{Appendix A.3. Cooperation under greedy evaluation}

\begin{figure}[H]
    \centering
    \includegraphics[width=0.8\textwidth]{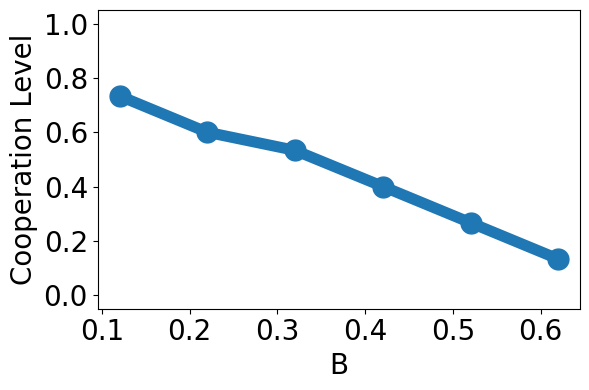}
    \caption{Average cooperation level as a function of the exploration strength $B$ at a fixed temptation parameter $D_r = 0.25$. For each value of $B$, results are averaged over 30 random seeds. All measurements are conducted during the evaluation phase with exploration disabled by setting $\tau_{\text{eval}} = 1$, such that agents act greedily with respect to their learned Q-functions.}
    \label{fig:greedy}  
\end{figure}

Figure~\ref{fig:greedy} reports the average cooperation level as a function of the exploration strength $B$ under a fixed temptation parameter $D_r = 0.25$. For each value of $B$, the reported cooperation level is obtained by averaging over 30 random seeds. Importantly, all measurements are conducted during the evaluation phase, in which exploration is disabled by setting $\tau_{\text{eval}} = 1$, such that agents act greedily with respect to their learned Q-functions.

As shown in the figure, cooperation decreases monotonically as $B$ increases, despite the absence of exploration noise during evaluation. At low exploration strengths, the learned policies sustain relatively high cooperation levels. However, as $B$ increases, cooperation progressively deteriorates and eventually collapses to low values, indicating that the adverse effect of strong exploration manifests through the learning dynamics rather than through stochastic action selection at test time.

This result demonstrates that the cooperation collapse observed in shared-policy DQN is not an artifact of exploratory behavior during evaluation. Instead, high exploration during training qualitatively alters the value representations learned by the agents, leading to policies that favor defection even when evaluated under fully greedy conditions. The robustness of this trend across 30 random seeds further suggests that the phenomenon reflects a systematic limitation of the learning architecture under strong exploration, rather than transient fluctuations or insufficient convergence.

\end{document}